\documentclass[prl,twocolumn,superscriptaddress,nobibnotes,amsmath,amssymb,showpacs]{revtex4}
\usepackage{graphicx}
\usepackage[utf8]{inputenc}
\begin{document}

\title{New scalings in nuclear fragmentation}

\author{E.~Bonnet}
\affiliation{GANIL, (DSM-CEA/CNRS/IN2P3), 
F-14076 Caen cedex, France}
\affiliation{Institut de Physique Nucl\'eaire, CNRS/IN2P3,
Universit\'e Paris-Sud 11, F-91406 Orsay cedex, France}
\author{B.~Borderie}
\affiliation{Institut de Physique Nucl\'eaire, CNRS/IN2P3,
Universit\'e Paris-Sud 11, F-91406 Orsay cedex, France}
\author{N.~Le Neindre}
\affiliation{Institut de Physique Nucl\'eaire, CNRS/IN2P3,
Universit\'e Paris-Sud 11, F-91406 Orsay cedex, France}
\affiliation{LPC, CNRS/IN2P3, Ensicaen,
Universit\'{e} de Caen, 
F-14050 Caen cedex, France}
\author{Ad.~R.~Raduta}
\affiliation{Institut de Physique Nucl\'eaire, CNRS/IN2P3,
Universit\'e Paris-Sud 11, F-91406 Orsay cedex, France}
\affiliation{National Institute for Physics and Nuclear Engineering,
RO-76900 Bucharest-Magurele, Romania}
\author{M.~F.~Rivet}
\affiliation{Institut de Physique Nucl\'eaire, CNRS/IN2P3,
Universit\'e Paris-Sud 11, F-91406 Orsay cedex, France}
\author{R.~Bougault}
\affiliation{LPC, CNRS/IN2P3, Ensicaen,
Universit\'{e} de Caen, 
F-14050 Caen cedex, France}
\author{A.~Chbihi}
\affiliation{GANIL, (DSM-CEA/CNRS/IN2P3), 
F-14076 Caen cedex, France}
\author{J.D.~Frankland}
\affiliation{GANIL, (DSM-CEA/CNRS/IN2P3), 
F-14076 Caen cedex, France}
\author{E.~Galichet}
\affiliation{Institut de Physique Nucl\'eaire, CNRS/IN2P3,
Universit\'e Paris-Sud 11, F-91406 Orsay cedex, France}
\affiliation{Conservatoire National des Arts et Métiers,
F-75141 Paris cedex 03, France}
\author{F.~Gagnon-Moisan}
\affiliation{Institut de Physique Nucl\'eaire, CNRS/IN2P3,
Universit\'e Paris-Sud 11, F-91406 Orsay cedex, France}
\affiliation{Laboratoire de Physique Nucléaire, Département de Physique,
de Génie Physique et d'Optique, Université Laval, Québec, Canada G1K 7P4}
\author{D.~Guinet}
\affiliation{Institut de Physique Nucl\'eaire, CNRS/IN2P3,
Universit\'e Claude Bernard Lyon 1, F-69622 Villeurbanne cedex, France}
\author{P.~Lautesse}
\affiliation{Institut de Physique Nucl\'eaire, CNRS/IN2P3,
Universit\'e Claude Bernard Lyon 1, F-69622 Villeurbanne cedex, France}
\author{J.~{\L}ukasik}
\affiliation{Institute of Nuclear Physics IFJ-PAN, PL-31342 Krak{\'o}w, Poland}
\author{P.~Marini}
\affiliation{Institut de Physique Nucl\'eaire, CNRS/IN2P3,
Universit\'e Paris-Sud 11, F-91406 Orsay cedex, France}
\affiliation{GANIL, (DSM-CEA/CNRS/IN2P3), 
F-14076 Caen cedex, France}
\author{M.~P\^arlog}
\affiliation{LPC, CNRS/IN2P3, Ensicaen,
Universit\'{e} de Caen, 
F-14050 Caen cedex, France}
\affiliation{National Institute for Physics and Nuclear Engineering,
RO-76900 Bucharest-Magurele, Romania} 
\author{E.~Rosato}
\affiliation{Dipartimento di Scienze Fisiche e Sezione INFN, Universit\`a
di Napoli ``Federico II'', I-80126 Napoli, Italy}
\author{R.~Roy}
\affiliation{Laboratoire de Physique Nucléaire, Département de Physique,
de Génie Physique et d'Optique, Université Laval, Québec, Canada G1K 7P4}
\author{G.~Spadaccini}
\affiliation{Dipartimento di Scienze Fisiche e Sezione INFN, Universit\`a
di Napoli ``Federico II'', I-80126 Napoli, Italy}
\author{M.~Vigilante}
\affiliation{Dipartimento di Scienze Fisiche e Sezione INFN, Universit\`a
di Napoli ``Federico II'', I-80126 Napoli, Italy}
\author{J.P.~Wieleczko}
\affiliation{GANIL, (DSM-CEA/CNRS/IN2P3), 
F-14076 Caen cedex, France}
\author{B.~Zwieglinski}
\affiliation{The Andrzej Soltan Institute for Nuclear Studies, PL-00681
Warsaw, Poland}
\collaboration{INDRA and ALADIN Collaborations} \noaffiliation

\begin{abstract}
Fragment partitions of fragmenting hot nuclei produced in central and
semiperipheral collisions have been compared in the excitation energy region
4-10 MeV per nucleon where radial collective expansion takes place.
It is shown that, for a  given total excitation energy per nucleon,
the amount of radial collective energy fixes the mean fragment multiplicity. 
It is also shown that, at a given total excitation
energy per nucleon, the different properties of fragment partitions are completely
determined by the reduced fragment multiplicity (fragment multiplicity
normalized to the source size). Freeze-out volumes seem to play a role
in the scalings observed.
\end{abstract}

\date{\today}
\maketitle

The process of the total disintegration of a nucleus accompanied by a copious
production of nuclear fragments, termed multifragmentation, was
discovered more than forty years ago~\cite{Gag70}.
However, tremendous progress in the understanding
of multifragmentation came
during the last fifteen years with the advent of powerful 4$\pi$
detectors~\cite{WCI06,Vio06,Bor08}.
From the theoretical side, the equation of state describing nuclear matter,
similar to the van der Waals equation for classical fluids, foresees the
existence of a liquid-gas type phase transition and multifragmentation
was long assimilated to this transition. In the last ten years the general theory
of first order phase transition in finite systems was strongly 
progressing~\cite{Gro01,Gul04,Cho04}.
Today a rather coherent picture
has been reached which is related to the experimental observation of different
phase transition signatures for hot multifragmenting
nuclei~\cite{Poc95,MDA00,I31-Bor01,Nato02,I40-Tab03,MDA04,Bor08,Bru08,I72-Bon09}.
The region of phase coexistence was identified in a semi-quantitative way
(3 to 10 AMeV excitation energy and 0.7 to 0.2 times the normal density)
and it is experimentally observed that fluctuations (configurational energy
and charge of the heaviest fragment) are large and even maximum in the middle
of this region for multifragmenting nuclei produced in semiperipheral
collisions~\cite{I63-NLN07}. Such large fluctuations seem to correlate points at a
distance comparable to the linear size of the system and can therefore have
an effect, in the coexistence region, similar to a diverging correlation length
in an infinite system at the critical point~\cite{I63-NLN07,Gul99,Yeo92}.
Consequently one can expect to reveal new
specific scalings related to a first order phase transition for finite systems.
To do that, in this Letter, we compare, with the help of the reduced
fragment multiplicity, the fragment partition properties
of fragmenting hot nuclei produced in central and semiperipheral collisions.

\begin{figure*}[htbp]
\begin{center}
\includegraphics*[scale=0.88]{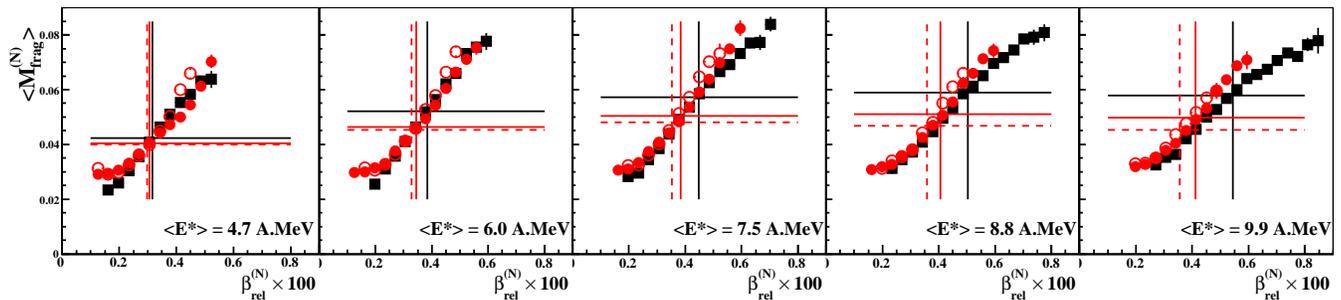}
\caption{(color online) 
Evolution of the average fragment
multiplicity normalized to the source charge/size $<M^{(N)}_{frag}>$=
$<M_{frag}/Z_s>$  as a function of the relative velocity of fragments,
$\beta^{N}_{rel}$, (see text) for different total excitation energy per
nucleon of the sources. Full squares, open and full circles stand 
respectively for QF sources and QP sources produced at 80 and 
100~MeV/nucleon incident energies. Crosses correspond to the mean values of the
considered samples for QF (black) and QP (grey dashed-80 MeV/nucleon and
full-100 MeV/nucleon) samples, see text.}
\label{fig1}
\end{center}
\end{figure*}
The experiments were performed with the $4\pi$ multidetector for
charged reaction products INDRA~\cite{I3-Pou95}. For central collisions, 
beams of $^{129}$Xe, accelerated by the GANIL accelerator in Caen, France, at five
incident energies: 25, 32, 39, 45 and 50 MeV/nucleon, bombarded a thin
target of natural tin (350~$\mu$g/$cm^2$) and hot quasi-fused (QF) 
nuclei/sources with Z around
70-100 could be selected~\cite{I40-Tab03}. Hot nuclei with lower Z around 60-70
(quasi-projectile, QP, sources) were obtained in
semi-peripheral Au+Au collisions at 80 and 100 MeV/nucleon incident energies at the
SIS heavy ion synchrotron at the GSI facility in Darmstadt, Germany~\cite{I45-Luk03}.
In this experiment the
$^{197}$Au beam was impinging on a 2 mg/cm$^{2}$ thick target. For both
experiments, Z identification was obtained over the whole range of fragments
produced
and the energy calibration was achieved with an accuracy of 4\%.
Further details can be found in~\cite{I14-Tab99,I34-Par02,I44-Trz03}.

The procedures used to select the studied nuclei/sources are the following.
First of all poorly-measured events were rejected by requiring the detection of
at least 80\% of the charge of
the initial system (projectile plus target for QF sources, projectile
only for QP sources). Then compact
sources were selected by using topology selectors in velocity space.
For QF sources the constraint of large flow angle ($\geq
60^{\circ}$)
calculated with the kinetic energy tensor for fragments (Z$\geq$5) in the
center of mass of the reaction was imposed (for details see~\cite{I40-Tab03}).
For QP sources it was done by imposing a
maximum value (2/3 of the QP velocity) of
the mean relative velocity between fragments emitted in the forward
hemisphere of the center
of mass (see Eqs.(1),(2) and~\cite{I69-Bon08} for details).
A constant QP size, within 10\%, was additionally required. Note that for QP
sources fission events were removed~\cite{I61-Pic06}.
Further information concerning selections and light charged particles
associated to sources can be found in~\cite{I69-Bon08}.
The following step consists in the evaluation of the total excitation energy of the 
different sources. The calorimetric method~\cite{Viol06} was used event by event.
Neutrons are not detected but their multiplicity is estimated from
the difference between the mass of the source and the sum of the masses
attributed to the detected
charged products. The source is assumed to have the same N/Z ratio as
the initial system.
\begin{table}
\begin{center}
\begin{tabular}{|c|c|c|c|c|c|}
\hline $E_{inc}$ (MeV/nucleon)		       		& 25 & 32 & 39 & 45 & 50  \\
\hline $<E^{*}>$ (A.MeV)	& 4.7 & 6.0 & 7.5 & 8.8 & 9.9 \\  
\hline $\sigma_{E^{*}}$ (A.MeV) 		      	& 0.7 & 0.8 & 1.0 & 1.1  & 1.3 \\  
\hline
\end{tabular}
\caption{ Selected excitation energy bins for the comparison between QF and
QP sources. The five columns stand for the five bombarding energies for
producing QF sources (fist line). 
Second and third lines indicate the mean value and root mean square of
excitation energy ($E^{*}$) distributions for QF sources. For comparisons,
QF and QP events are limited to those with
excitation energies defined as $|E^{*}-<E^{*}>| \le 0.3\;\sigma_{E^{*}}$.}
\label{tab1}
\end{center}
\end{table}
Hypotheses which have been made
for QF sources are the following: a level density
parameter equal to A/10, the average kinetic energy of neutrons equal to their
emitting source temperature and the Evaporation Attractor Line formula 
(A=Z(2.072+2.32$\times10^{-3}$Z))~\cite{Cha98} used to calculate fragment
masses. EAL is especially well-adapted when heavy fragments
($Z>20$) result from the deexcitation of neutron deficient sources. 
For QP sources the hypotheses are identical except
for calculated fragment masses for which 
we use the formula (A=Z(2.045+3.57$\times10^{-3}$Z))~\cite{Cha98}, 
better adapted for excited nuclei close to the beta-stability valley. 
Note that, compared to the EAL
formula, differences appear only for masses associated to Z greater than 40.

In the following, we will use as main sorting parameter the total 
excitation energy per nucleon, $E^*$, to compare fragment
properties of both QF and QP sources.
Excitation energy bins of compared
samples have been fixed by values (mean and $\sigma$) calculated for QF sources at the five
incident energies (see Table I). Only the mean values are reported in the
panels of figures.

We suppose that fragment
velocities, in their source frame, are the results of the composition of three components: a
randomly-directed thermal contribution, a Coulomb contribution
dependent on the fragment charges and source sizes, and a radial collective energy.
To evaluate the radial collective energy involved in the de-excitation of the
different sources the mean relative velocity between fragments
\begin{eqnarray}
\mathrm\;\beta_{rel}=\frac{2}{M_{frag}(M_{frag}-1)}\sum_{i < j}|
\vec{\beta^{(ij)}}|\\
\qquad \vec{\beta^{(ij)}}=\vec{\beta^{(i)}}-\vec{\beta^{(j)}}
\end{eqnarray}
is used, which is independent of the reference frame. Only events or 
subevents (for QP sources)
with fragment multiplicities $M_{frag}$ greater than one are considered. 
It was shown in~\cite{I69-Bon08} that this observable is a good measure
of the amount of radial collective energy. 
The thermal component of fragment velocities has a negligible contribution
to their mean relative velocity, while increasing the size of fluctuations.
The effect of the
Coulomb contribution is removed by using a simple normalization 
\begin{eqnarray}
\label{eq_vrn} \beta^{(N)}_{rel} = \frac{\beta_{rel}}{\sqrt{<Z>(Z_{s}-<Z>)}}
\end{eqnarray}
which takes
into account, event by event, the Coulomb influence of
the mean fragment charge ($<Z>$) on the complement of the source charge
($Z_s-<Z>$). 

We first look at the most global observable: the mean fragment multiplicity.
Figure 1 shows that, for a given total excitation energy per nucleon, the mean fragment
multiplicity normalized to the source charge/size $<M^{(N)}_{frag}>$=
$<M_{frag}/Z_s>$ is strongly correlated with the amount of radial collective
expansion, as measured by $\beta^{N}_{rel}$.
Crosses correspond to mean values for each type of source.
Depending on the source type the
relative contributions of radial collective energy (thermal pressure
and compression-expansion cycle) strongly differ~\cite{I69-Bon08}
but in all cases $\beta^{N}_{rel}$,
representative of the total collective energy, fixes the normalized mean
fragment multiplicity.
Note that for the largest $\beta^{N}_{rel}$
values (above 0.006) reached only at the higher excitation energies for QF
sources, normalized
fragment multiplicities increase more slowly. In that case the average size
of fragments decreases and a larger fraction of them have a charge below
our fragment limit Z=5.
\begin{figure}[htbp]
\begin{center}
\resizebox{\columnwidth}{!}{\includegraphics{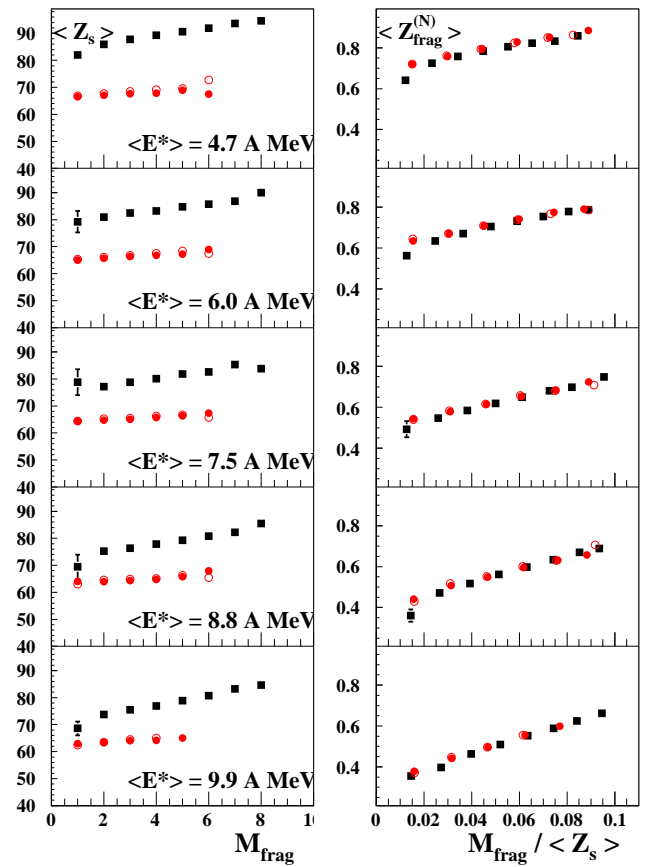}}
\caption{(color online) 
Left side refers to mean values of source charges as a function of fragment
multiplicity for different total excitation energy per nucleon of the
sources. Right side shows the evolution of the total charge bound in
fragments normalized to the source charge/size, $Z_{frag}^{(N)}$, as 
a function of the reduced fragment multiplicity $M_{frag}$/$<Z_s>$ 
for the same total excitation energies. Full squares, open and full 
circles stand respectively for QF sources and QP
sources produced at 80 and 100 MeV/nucleon incident energies.}
\label{fig2}
\end{center}
\end{figure}
\begin{figure}[htbp]
\begin{center}
\resizebox{\columnwidth}{!}{\includegraphics{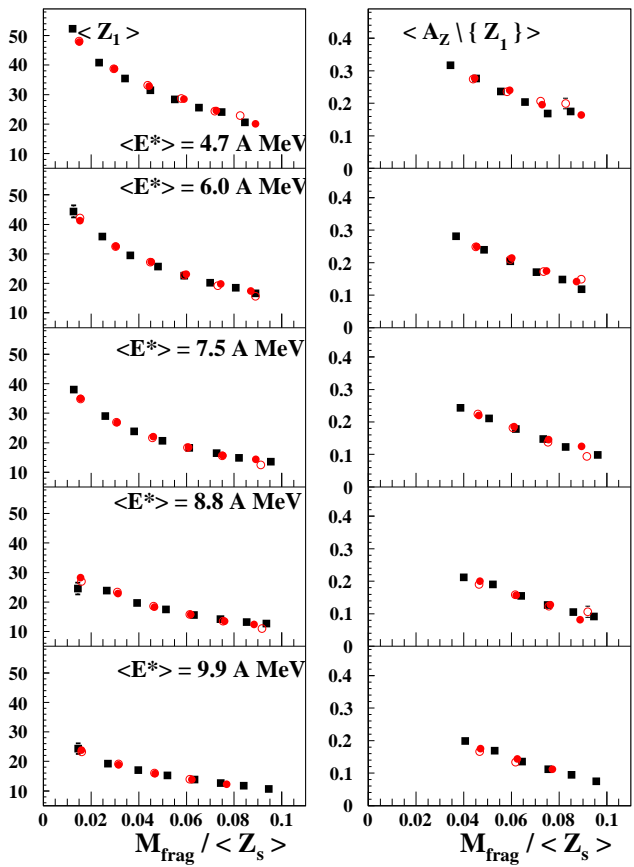}}
\caption{(color online) 
Left and right sides refer respectively to the mean charge of the
heaviest fragment of partitions, $<Z_1>$,
and to the generalized asymmetry in charge of the fragment partitions without
the heaviest one, $A_{Z} \backslash \{Z_{1}\}$, (see text) as a function of 
the reduced fragment
mutiplicity, $M_{frag}$/$<Z_s>$, for different total excitation energy per
nucleon of the sources. Full squares, open and full circles stand respectively 
for QF sources and QP sources produced at 80 and 100 MeV/nucleon incident 
energies.}
\label{fig3}
\end{center}
\end{figure}

One can now go a step further and study the properties of fragment partitions
(total charge bound in fragments, heaviest fragment, asymmetry in charge of
fragments) associated
to the different values of the fragment multiplicity distributions.
To do that, and to be able to compare QF and QP sources, we introduce a new variable the
reduced fragment multiplicity $M_{frag}$/$<Z_s>$. It is built from Fig. 2
(left side) which shows the evolutions of the mean source size with the
different values, $M_{frag}$, of the fragment multiplicity distributions
for the different excitation energies per nucleon.
Figure 2 (right side) shows the evolution with the reduced fragment
multiplicity of the total charge bound in
fragments normalized to the source charge/size, $Z_{frag}^{(N)}$, for the different
excitation energies per nucleon. We observe that the mean values of $Z_{frag}^{(N)}$
are independent of the source type.
On the left side of Fig.3
average values of the heaviest fragment charge, $Z_1$, are reported: values for QP and QF
sources follow exactly the same evolution. The independence of $Z_1$ on the
system size was already observed and is valid as long as the total system or
projectile mass is above 190~\cite{I12-Riv98,Bor08}.
Finally the division of
the charge among the other fragments is investigated using the generalized charge asymmetry
of the fragment partitions,
\begin{eqnarray}
\label{eq_az} A_{Z}= \frac{1}{\sqrt{M_{frag}-1}}\frac{\sigma_{Z}}{<Z>}
\end{eqnarray}
which was introduced in~\cite{I69-Bon08}.
One can re-calculate the generalized asymmetry by removing
$Z_{1}$ from partitions with at least 3 fragments noted 
$A_{Z} \backslash \{Z_{1}\}$. The results are displayed in Fig. 3 (right
side). They follow a linear behavior and do not depend on the source type at a given
total excitation energy per nucleon.

To summarize,
we have shown in this Letter new results independent of the mechanism of
production of hot fragmenting nuclei. First, at a given total excitation energy per nucleon,
the amount of the radial collective energy fixes the mean fragment
multiplicity normalized to the charge/size of the fragmenting hot
nucleus; we recall that
in the multifragmentation regime (3-10 AMeV) the radial
collective energy is limited to at most 20\%
of the total excitation energies involved~\cite{I69-Bon08}.
And secondly, the
properties of fragment partitions are completely determined by the reduced
fragment multiplicity. 
What is the possible origin of
the scalings observed? To, at least, give a direction to follow for future
studies, we have checked to what extent fragment partition scalings are
compatible with statistical multifragmentation models with cluster degrees
of freedom. The standard version of the Microcanonical Multifragmentation Model
(MMM)~\cite{Rad97,Rad02} was used. Inputs were the average A and Z of QF and QP sources
at the different total excitation energies per nucleon.
Contributions of radial collective energy to the total excitation energy were
taken from~\cite{I69-Bon08}. 
We do obtain fragment partition scalings if and only if the freeze-out
volume of the QF sources increases monotonically with the total excitation
energy from 2.7 to 5.7 $V_0$  and the freeze-out volume of the QP sources is
kept fixed at the intermediate value of 3.5 $V_0$; $V_0$ would correspond to
the volume of the source at normal density. However the agreement between MMM results
and the present observed
scalings is partial: a rather fair agreement for $Z_{frag}^{(N)}$
and $Z_1$ and a systematic underestimation for the fragment asymmetry without
$Z_1$. We recall that in MMM the radial collective energy is decoupled from
the break-up thermodynamics and fragment formation.
Clearly dynamical approaches, which
contain the coupling between thermal and collective expansion effects to
produce fragments~\cite{Riz07}, must be compared to those data.

%
%

\end{document}